\documentclass[floatfix, noshowpacs, preprintnumbers, twocolumn, amsmath, amssymb, aps, superscriptaddress, prb]{revtex4}

\pdfoutput=1

\usepackage{graphicx, xcolor}
\usepackage[caption=false]{subfig}
\usepackage{soul}

\newcommand{\ket}[1]{\left| #1 \right>}

\begin{document}

\title{Experimental Quantum Fast Hitting on Hexagonal Graphs}

\author{Hao Tang}
\affiliation{State Key Laboratory of Advanced Optical Communication Systems and Networks, Institute of Natural Sciences $\&$ Department of Physics and Astronomy, Shanghai Jiao Tong University, Shanghai 200240, China}
\affiliation{Synergetic Innovation Center of Quantum Information and Quantum Physics, University of Science and Technology of China, Hefei, Anhui 230026, China}
\author{Carlo Di Franco}
\affiliation{School of Physical and Mathematical Sciences, Nanyang Technological University, 637371, Singapore}
\affiliation{Complexity Institute, Nanyang Technological University, 637723, Singapore}
\author{Zi-Yu Shi}
\affiliation{State Key Laboratory of Advanced Optical Communication Systems and Networks, Institute of Natural Sciences $\&$ Department of Physics and Astronomy, Shanghai Jiao Tong University, Shanghai 200240, China}
\affiliation{Synergetic Innovation Center of Quantum Information and Quantum Physics, University of Science and Technology of China, Hefei, Anhui 230026, China}
\author{Tian-Shen He}
\affiliation{State Key Laboratory of Advanced Optical Communication Systems and Networks, Institute of Natural Sciences $\&$ Department of Physics and Astronomy, Shanghai Jiao Tong University, Shanghai 200240, China}
\author{Zhen Feng}
\affiliation{State Key Laboratory of Advanced Optical Communication Systems and Networks, Institute of Natural Sciences $\&$ Department of Physics and Astronomy, Shanghai Jiao Tong University, Shanghai 200240, China}
\affiliation{Synergetic Innovation Center of Quantum Information and Quantum Physics, University of Science and Technology of China, Hefei, Anhui 230026, China}
\author{Jun Gao}
\affiliation{State Key Laboratory of Advanced Optical Communication Systems and Networks, Institute of Natural Sciences $\&$ Department of Physics and Astronomy, Shanghai Jiao Tong University, Shanghai 200240, China}
\affiliation{Synergetic Innovation Center of Quantum Information and Quantum Physics, University of Science and Technology of China, Hefei, Anhui 230026, China}
\author{Ke Sun}
\affiliation{State Key Laboratory of Advanced Optical Communication Systems and Networks, Institute of Natural Sciences $\&$ Department of Physics and Astronomy, Shanghai Jiao Tong University, Shanghai 200240, China}
\author{Zhan-Ming Li}
\affiliation{State Key Laboratory of Advanced Optical Communication Systems and Networks, Institute of Natural Sciences $\&$ Department of Physics and Astronomy, Shanghai Jiao Tong University, Shanghai 200240, China}
\affiliation{Synergetic Innovation Center of Quantum Information and Quantum Physics, University of Science and Technology of China, Hefei, Anhui 230026, China}
\author{Zhi-Qiang Jiao}
\affiliation{State Key Laboratory of Advanced Optical Communication Systems and Networks, Institute of Natural Sciences $\&$ Department of Physics and Astronomy, Shanghai Jiao Tong University, Shanghai 200240, China}
\affiliation{Synergetic Innovation Center of Quantum Information and Quantum Physics, University of Science and Technology of China, Hefei, Anhui 230026, China}
\author{Tian-Yu Wang}
\affiliation{State Key Laboratory of Advanced Optical Communication Systems and Networks, Institute of Natural Sciences $\&$ Department of Physics and Astronomy, Shanghai Jiao Tong University, Shanghai 200240, China}
\author{ M. S. Kim}
\affiliation{QOLS, Blackett Laboratory, Imperial College London, SW7 2AZ, UK}
\author{Xian-Min Jin}
\email{xianmin.jin@sjtu.edu.cn} 
\affiliation{State Key Laboratory of Advanced Optical Communication Systems and Networks, Institute of Natural Sciences $\&$ Department of Physics and Astronomy, Shanghai Jiao Tong University, Shanghai 200240, China}
\affiliation{Synergetic Innovation Center of Quantum Information and Quantum Physics, University of Science and Technology of China, Hefei, Anhui 230026, China}

\maketitle
\textbf{Quantum walks are powerful kernels in quantum computing protocols that possess strong capabilities in speeding up various simulation and optimisation tasks. One striking example is given by quantum walkers evolving on glued trees for their faster hitting performances than in the case of classical random walks. However, its experimental implementation is challenging as it involves highly complex arrangements of exponentially increasing number of nodes. Here we propose an alternative structure with a polynomially increasing number of nodes. We successfully map such graphs on quantum photonic chips using femtosecond laser direct writing techniques in a geometrically scalable fashion. We experimentally demonstrate quantum fast hitting by implementing two-dimensional quantum walks on these graphs with up to 160 nodes and a depth of 8 layers, achieving a linear relationship between the optimal hitting time and the network depth. Our results open up a scalable way towards quantum speed-up in complex problems classically intractable. }

Adapting well-known classical mathematical models in a way to include quantum mechanical laws has shown the emergence of new interesting behaviors. In some cases, the modified protocols have revealed an advantage with respect to the original ones in solving specific problems. This has clearly triggered the interest of the scientific community in the quest for a better understanding and exploitation of these new tools \cite{Mohseni2017}. A striking example is given by quantum walks, the adaptation of the classical random walk to the world of quantum mechanics\cite{Childs2008}. Quantum walks have already found applications in several scenarios, including spatial search problems\cite{Szegedy2004,Childs2004}, the element distinctness problem\cite{Aaronson2004}, testing matrix identities\cite{Buhrman2006}, evaluating Boolean formulas\cite{Farhi2008}, judging graph isomorphism\cite{Douglas2008,Bruderer2016}, which all theoretically promise quantum speed-up and may inspire the breakthrough in real-life applications.

\begin{figure*}[ht!]
\includegraphics[width=0.79\textwidth]{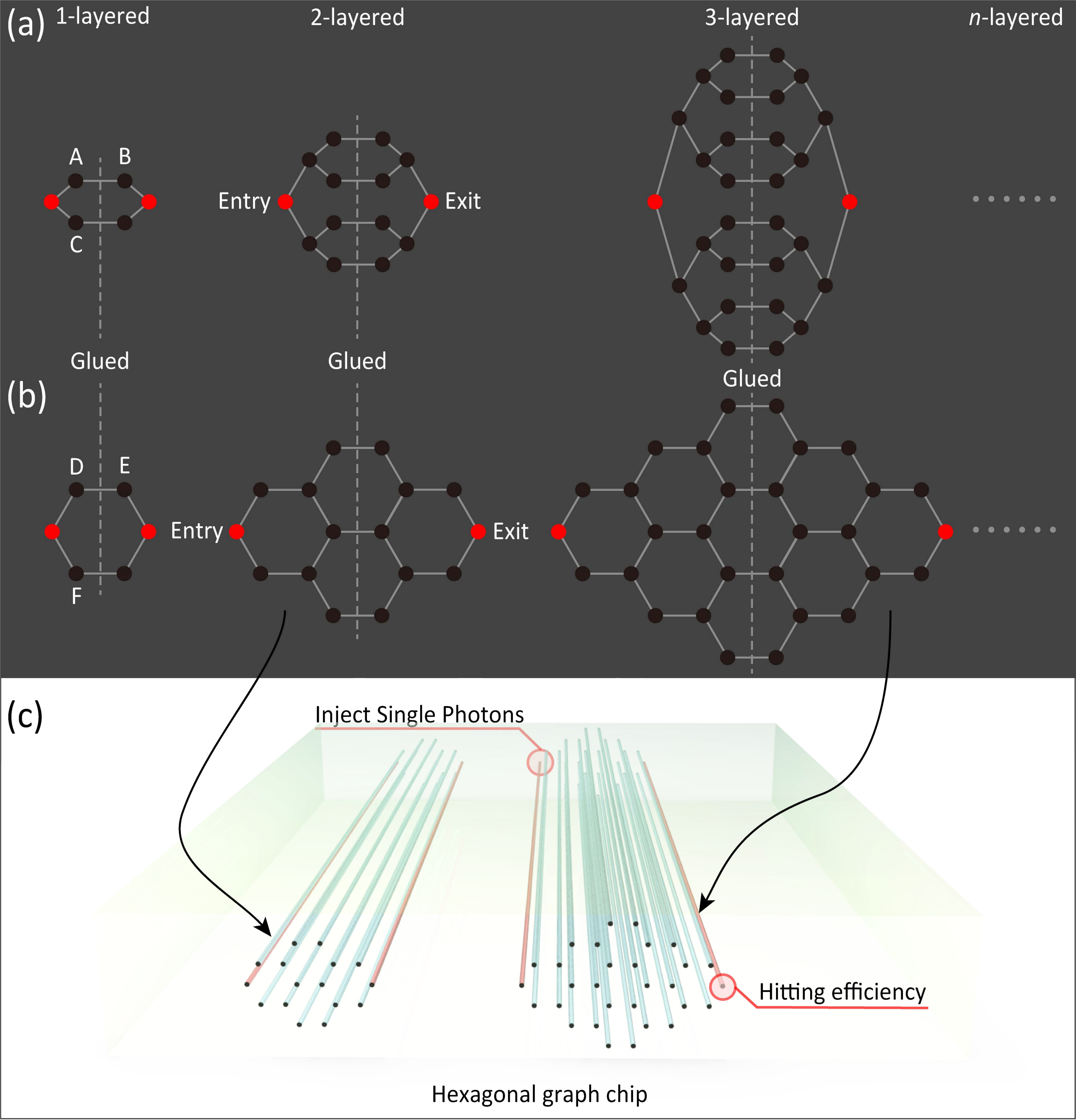}
\caption{\textbf{Theoretical graphs and their implementation on photonic chips.} Schematic diagram of ({\bf a})  a binary glued tree and ({\bf b})
the proposed hexagonal graph. ({\bf c}) Schematic diagram of quantum fast-hitting experiment on hexagonal graphs based on femtosecond laser written waveguide arrays.}
\label{fig:QFTConcept}
\end{figure*}

One feature of quantum walks on complex graphs that is key in quantum algorithms is their ability to propagate from a node to a distant one in an efficient way. This is often denoted as fast hitting. In particular, fast hitting on a structure known as glued tree is extremely charming due to its exponential speed-up over its classical counterpart\cite{Childs2002, Childs2003}. A glued tree is obtained by connecting the ``final leaves'' of two binary tree graphs\cite{Farhi1998} of the same depth, as shown in Fig.\ref{fig:QFTConcept}{\bf(a)}. The  process assumes a particle starting in the left-most vertex (called the Entry site), evolving through the graph, and finally hitting the right-most vertex (called the Exit site).
It has been shown that, in a scenario where the central connections are randomly chosen, any algorithm exploiting a classical walker ({\it i.e.}, a particle following the laws of classical mechanics) would require on average a time scaling exponentially with the graph depth to reach the Exit. On the other hand, a quantum walker will require a time that scales only linearly \cite{Childs2002,Douglas2009,Carneiro2005}. Due to the close relation between binary trees and decision trees in computer science, this could generate enormous benefits if properly incorporated into real optimization problems. 

Unfortunately, an implementation of quantum walks on this class of graphs is not feasible with the current technology. The fact that the number of vertices grows exponentially with the size of the graph itself is one of the main hurdles for their realization. However, even showing the speed-up by a quantum walker over a classical walker on a simpler graph (where, for instance, the number of vertices grows quadratically), is already of great interest: this would be a pioneering experimental demonstration of the quantum advantage in algorithms based on quantum walks on tree structures. So far, one-dimensional quantum walks have been successfully realized in various physical systems\cite{Schmitz2009,Karski2009,Broome2010,Sansoni2012,Cardano2015,Du2003,Perets2008, Peruzzo2010,Preiss2015}, and two-dimensional quantum walks have been demonstrated with time-polarization degrees of freedom\cite{Schreiber2012, Jeong2013} or genuine spatial two-dimensional structures\cite{Tang2017}. However, an experimental demonstration of the hitting time advantage given by quantum walks on complex structures has never been shown yet.

Here we present a modified version of the glued tree structure that can be mapped into a photonic chip and keep excellent extendibility for larger complexity. We study the hitting efficiency against the evolution time when increasing the network layer depth, representing the network complexity, going from 2 to 8. We demonstrate that the time for optimal hitting increases linearly with the layer depth, giving a quadratic speed-up over its classical counterparts.  

\begin{figure*}[ht!]
\includegraphics[width=0.9\textwidth]{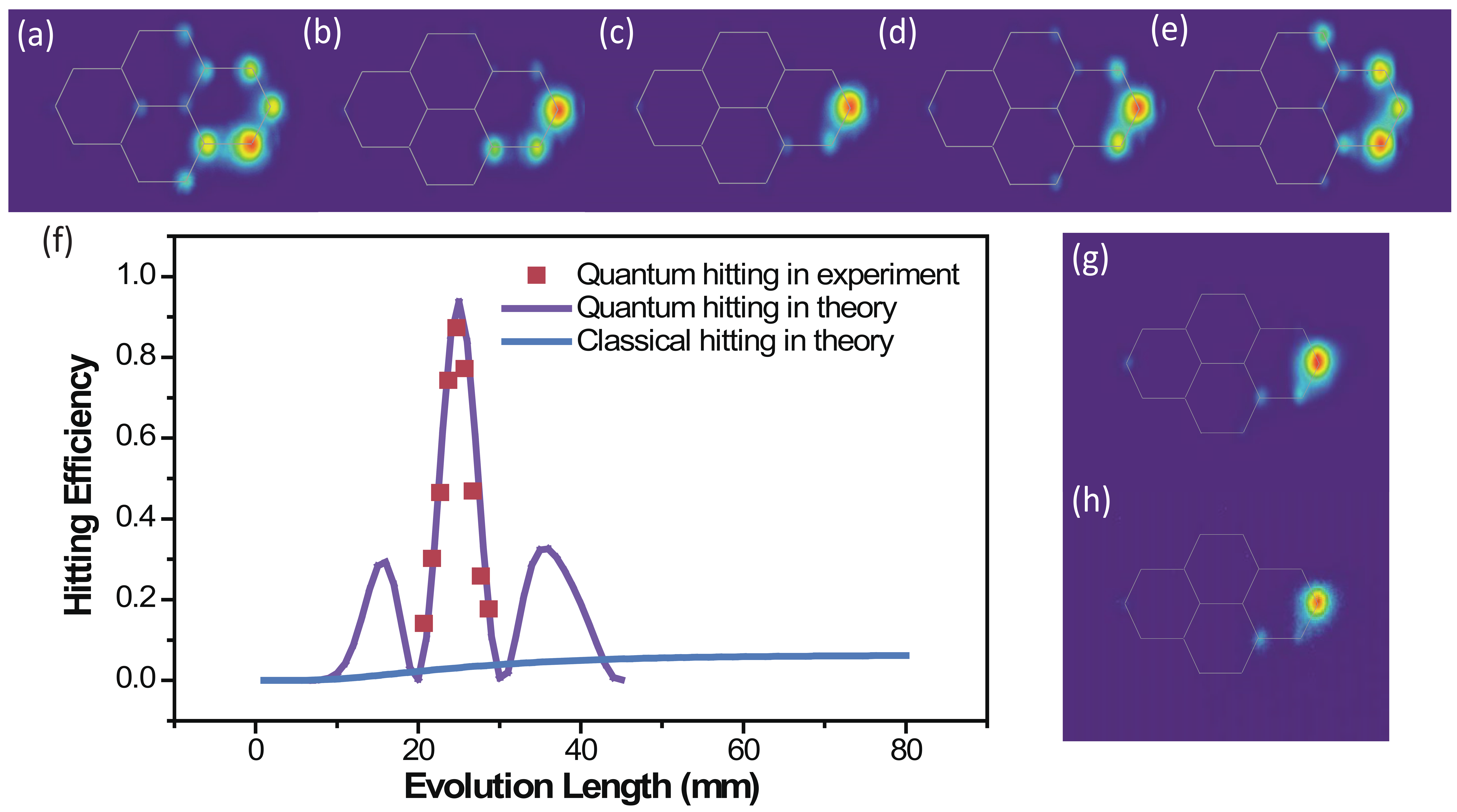}
\caption{\textbf{Fast hitting on a 2-layered hexagonal graph.} ({\bf a-e}) Photographed evolution patterns for a 2-layered hexagonal graph at different evolution lengths: ({\bf a}) 20.7mm, ({\bf b}) 22.7mm, ({\bf c}) 24.7mm, ({\bf d}) 26.7mm, and ({\bf e}) 28.7mm. ({\bf f})  The hitting efficiency against the evolution length for quantum hitting and classical hitting. The evolution patterns for the same sample with an evolution length equal to 25.2mm, by injecting ({\bf g}) laser beam and ({\bf h}) heralded single photons, respectively.}
\label{fig:strutturaChip}
\end{figure*}

As is shown in Fig.\ref{fig:QFTConcept}{\bf(a)} and Fig.\ref{fig:QFTConcept}{\bf(b)}, the hexagonal structure proposed here resembles the binary glued tree as they are both obtained by gluing two tree-like structures. In our mapping into a three-dimensional waveguide scheme [shown in Fig.\ref{fig:QFTConcept}{\bf(c)}], the cross section of the three-dimensional array corresponds to the desired graph, with each waveguide representing a node, while the longitudinal propagation direction corresponds to the evolution time. 

When photons are injected into the Entry site, they propagate along this waveguide, and meanwhile evanescently couple to other waveguides\cite{Gao2016,Tang2017}. As the coupling coefficient decays exponentially when the waveguide spacing increases \cite{Szameit2007}, only the coupling between the most adjacent waveguides are considered here, representing a connected path in the graph [{\it e.g.}, Site A-B in Fig.\ref{fig:QFTConcept}{\bf(a)}, Site D-E in Fig.\ref{fig:QFTConcept}{\bf(b)}]. Waveguides further apart can be considered disconnected due to the marginal coupling coefficient (e.g. Site A-C, Site D-F). In the hexagonal structure, the layer depth corresponds to the number of hexagons in the central column, as shown in Fig.1(b). When the layer depth increases, having an exponentially increasing number of waveguides disconnected in the photonic chips for the binary glued trees is clearly not feasible. On the other hand, we use the hexagonal structure that grows in a regular way and is possible to map on a photonic chip.

For a quantum walk that evolves along the waveguides, the propagation length $z$ is proportional to the propagation time $t$ according to $z = vt$, where $v$ is the speed of light in the waveguide, and hence all the terms that are a function of $t$ would use $z$ instead in this paper for simplicity. The initial wavefunction $\ket{\Psi(0)}$ evolves according to $$\ket{\Psi(z)}=e^{-iHz}\ket{\Psi(0)},\eqno{(1)}$$ where $H$ is the Hamiltonian that contains the information on the couplings of the photonic network. The evolved wavefunction can be obtained by matrix exponential methods when $H$ is known\cite{Izaac2015,Tang2017}. For the classical counterpart, we use the versatile quantum stochastic walk model\cite{Whitfield2010} and set it to the purely classical domain without any quantum term. We obtain the continuous-time dynamics for classical walks that can be compared to our continuous-time quantum walks. More theoretical details can be found in the Supplementary Note 1. 

It is worth noting that quantum walks intrinsically yield non-stationary solutions\cite{Sanchez2012}, which means that there exists an optimal hitting scenario at a certain evolution length. We would take notes of the optimal hitting efficiency and its corresponding optimal evolution length $z_o$. 

In light of theoretical predictions, we use femtosecond laser direct writing techniques\cite{Szameit2007,Crespi2013,Chaboyer2015,Feng2016} to fabricate 7 sets of hexagonal graphs with their layer depth varying from 2 to 8. For each layer depth $i$, we prepare 9 samples with their evolution length varying from $z_{oi}$-4mm to $z_{oi}$+4mm with intervals of 1mm, where $z_{oi}$ is the calculated optimal length for this structure. We characterize all samples and identify $z_{oi}$ for each set of graphs by injecting a vertically polarized 780nm laser beam into the Entry site and capturing the evolution pattern with a CCD camera (see Supplementary Note 2 and 3). With heralded single photons, we then directly observe the evolution pattern of the characterized $z_{oi}$ (see Methods for the details of single-photon generation and measurement).

\begin{figure*}[ht!]
\includegraphics[width=0.78\textwidth]{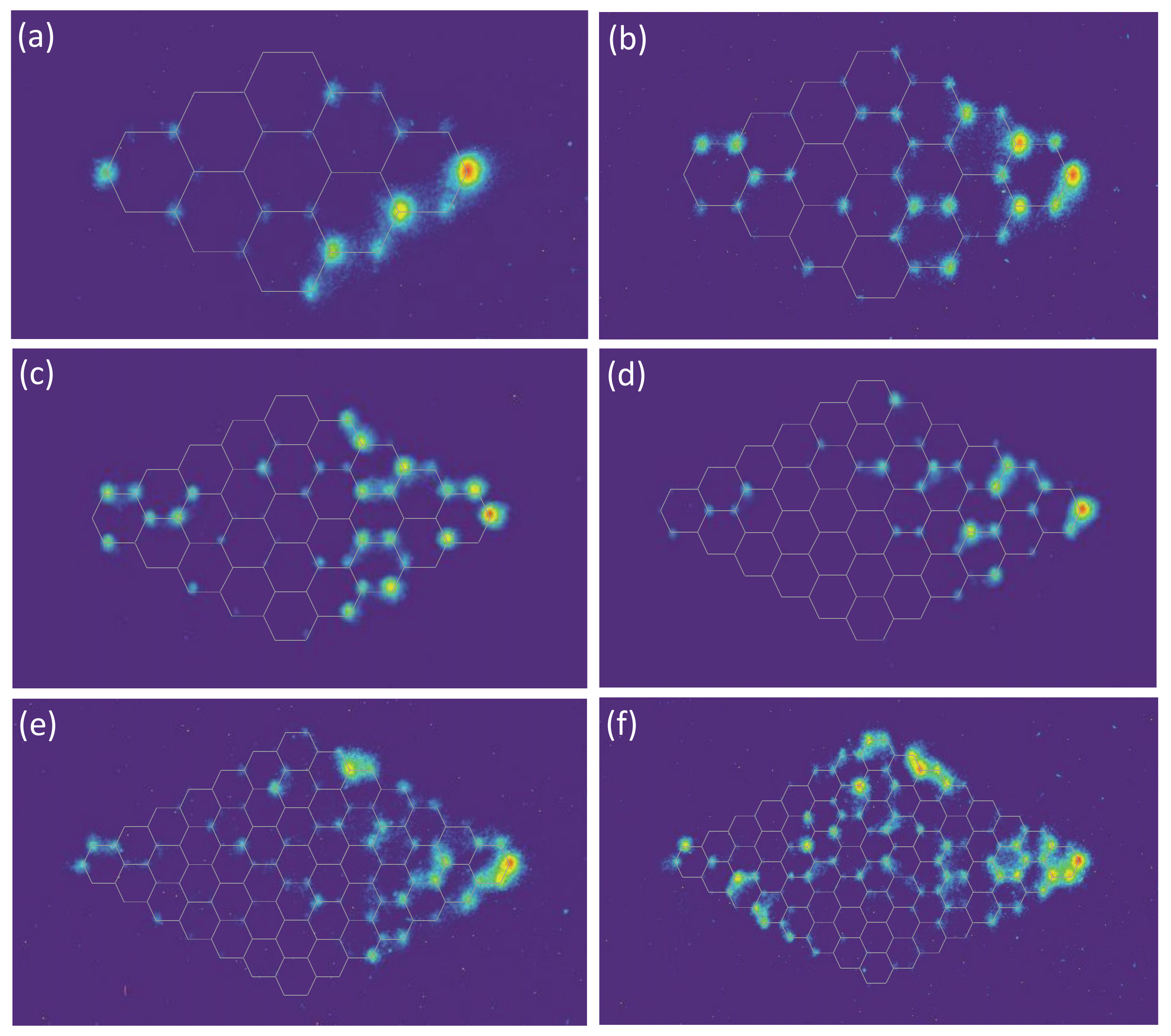}
\caption{\textbf{Increasing the complexity of hexagonal graphs.} ({\bf a-f}) Photographed evolution patterns for hexagonal graphs of different layer depths from 3 to 8. Each panel shows the optimal hitting scenario among the nine samples of the same layer depth. The single-photon-level imaged evolution patterns for: ({\bf a}) 3-layered graph at an evolution length of 30.4mm, ({\bf b}) 4-layered graph at 43.7mm, ({\bf c}) 5-layered graph  at 48.4mm, ({\bf d}) 6-layered graph  at 61.8mm, ({\bf e}) 7-layered graph  at 70.8mm, ({\bf f}) 8-layered graph  at 85.8mm. The experimental dynamics of hitting efficiency against the evolution time agrees well with our simulation.}
\label{fig:apparato}
\end{figure*}

From the photographed patterns for 2-layered hexagonal graphs in Fig.\ref{fig:strutturaChip}{\bf(a-e)}, the brightest spot at the Exit site occurs in Fig.\ref{fig:strutturaChip}{\bf(c)}, corresponding to the optimal hitting efficiency among these figures. The hitting efficiency against $z$ from the experiments agrees very well with the theoretical results, as shown in
Fig.\ref{fig:strutturaChip}{\bf(f)}, in terms of both the value of optimal hitting efficiency (of around 90\%) and the evolution length at which the optimal hitting occurs (at around 25mm). In order to show the advantage given by quantum walks, a comparison is made by adding in Fig.\ref{fig:strutturaChip}{\bf(f)} the theoretical result of the classical counterpart on the same structure: the classical hitting efficiency of only 6.25\% is outperformed by more than one order of magnitude. The low efficiency for the classic hitting is a result of the diffusive nature of the classical walk, and the optimal hitting efficiency is in fact the inverse of the total number of sites. On the other hand, the impressive advantage in the quantum case for the hitting task in networks with many binary paths comes from the interference governing the quantum evolution of the walker. We further polish the chip to seek even higher hitting efficiency suggested by the fitting result. The measured optimal evolution pattern at $z$ of 25.2 mm with both laser beam and heralded single photons are shown in Fig.\ref{fig:strutturaChip}{\bf(g-h)}. Our results confirmed that for one walker single photons and laser beam produce very consistent results. Furthermore, the implementation with genuine single photons represent a substantial step forward to a faithful realization of quantum fast hitting.

Extending the size of the hexagonal structure to a layer depth up to 8, we are able to show the quantum advantage in fast hitting for graphs of higher complexity. The evolution patterns with an optimal hitting efficiency out of the nine samples for each structure are measured using the heralded single photons, see Fig.\ref{fig:apparato}{\bf(a-f)}. We can see that the Exit site attracts more light than the others, and at least 50\% of the sites have barely any probability, a situation very different from the even distribution in the classical hitting. 

\begin{figure*}[ht!]
\includegraphics[width=1.0\textwidth]{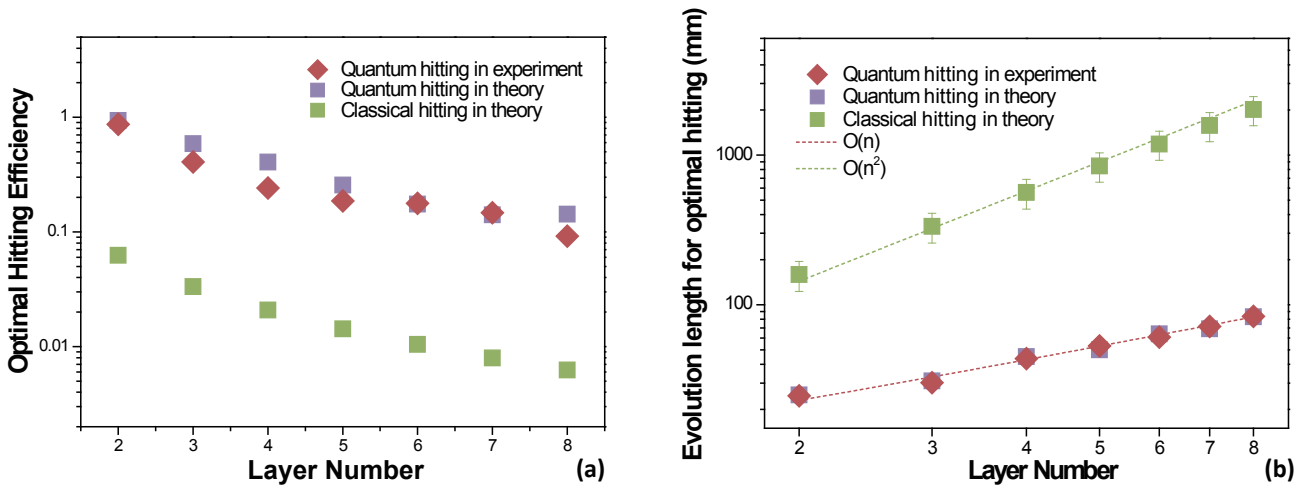}
\caption{\textbf{Comparison between quantum hitting and classical hitting}. ({\bf a}) Optimal hitting efficiency and ({\bf b}) the evolution length at which the optimal hitting occurs for hexagonal graphs of different layer depths.}
\label{fig:Results4}
\end{figure*}

We then analyze the performance of quantum and classical hitting on the hexagonal structures as a function of the layer depth. The optimal quantum hitting efficiency from both experiments and theory always remains more than one order of magnitude larger than the classical case, as shown in Fig.\ref{fig:Results4}{\bf(a)}. In this hexagonal structure, the total number of sites is $2n^2+4n$, where $n$ is the layer depth. Hence, the classical optimal hitting efficiency scales as $n^{-2}$. Fig.\ref{fig:Results4}{\bf(b)} presents the evolution length at which the optimal hitting efficiency occurs. The panel shows a clear linear trend for the quantum scenario, while its classical counterpart requires a quadratically larger evolution length.   

In conclusion, we have demonstrated the quantum fast hitting on hexagonal structures in photonic chips and experimentally observed that the time for optimal quantum hitting increases linearly with the layer depth. In comparison, the classical scenario is characterized by a quadratic relation; our investigation is therefore a demostration of the speed-up given by the interference that governs the evolution of the quantum walker, a key point in many tasks based on quantum walks. Overall there is a very good agreement between the experimental data and the theoretical predictions, for both the optimal efficiency and its corresponding evolution length. This work was made possible through the precise and versatile techniques of fabricating three-dimensional integrated photonic chips using femtosecond laser direct writing. Such capability paves the way for a broader and useful exploitation of quantum walks on complex graphs. 

In the future, it would be interesting to investigate the role of defects, asymmetry and photon bunching effect\cite{Gao2016} for quantum fast hitting. Experimental demonstration of the quantum advantage in fast hitting can also be extended to other graphs that have been extensively studied in theory, {\it e.g.} the hypercube graph, which also shows an exponential speed-up in quantum walks hitting\cite{Childs2008}. Finally, as binary trees are closely related to decision trees in computer science, we may utilize the quantum speed-up to improve the performance for tasks such as optimization, management and information searching. 

\section*{Methods}
\textbf{Fabrication of the hexagonal graph chip:}
The three-dimensional layout of hexagonal structure is designed according to the characterized coupling coefficients and programed by inputting the (x,y) coordinates for each site and the evolution length $z$ identically for all sites in each sample.
We feed a femtosecond laser (10W, 1026nm, 290fs pulse duration, 1MHz repetition rate and 513nm working wavelength) into an spatial light modulator (SLM) to create burst trains onto a borosilicate substrate with a 50X objective lens (numerical aperture of 0.55) at a constant velocity of 10mm/s. The waveguide arrays are prepared in a hexagonal structure by mapping it into the cross-section of the chip and the designed evolution length determines the array length. Power and SLM compensation are used to improve the uniformity\cite{Feng2016}.\\

\textbf{Heralded single-photon preparation and single-photon-level imaging:} Photon pairs with a wavelength of 780nm are efficiently generated via spontaneous parametric down conversion (SPDC)\cite{Kim2003,Kim2003b}. The pairs then go through a band-pass filter and a polarized beam splitter (PBS) to be divided into two components, horizontal and vertical polarization. We inject the vertically polarized photon into the Entry waveguide in the photonic chip, while the horizontally polarized photon is connected to a single photon detector that sets a trigger for heralding the horizontally polarized photons on a ICCD camera with a time slot of 10 ns. Without the external trigger, the measured patterns would come from the thermal-state light rather than single-photons. The ICCD camera captures each evolution pattern with a certain evolution length, after accumulating in the `external' mode for 1-1.5 hours.\\

\textbf{Hitting efficiency acquisition:} When collecting the data from experiments, we obtain the corresponding ASCII file, which is essentially a matrix of pixels. We create a `mask'  that contains the pixel coordinate of the circle centre and the
radius in pixels for each waveguide, and sum up the light intensity for all the pixels within each circle using Matlab. The normalized proportion of light intensity for each circle represents the probability at the corresponding waveguide. The hitting efficiency is the proportion of light intensity at the Exit site.\\

\section*{Acknowledgements} 
The authors thank J. D. Whitfield for a very useful conversation on numerical methods for the quantum stochastic walks, and J.-W. Pan for helpful discussions. This research is supported by the National Natural Science Foundation of China (Grant No. 11690033, 11374211, 11275131, 11571313, 11675113), the Innovation Program of Shanghai Municipal Education Commission (No. 14ZZ020), Shanghai Science and Technology Development Funds (No. 15QA1402200), and the open fund from HPCL (No. 201511-01). C. Di Franco is funded by the Singapore National Research Foundation Fellowship NRF-NRFF2016-02. M. S. Kim is supported by the Samsung GRO project, the EPSRC (EP/K034480/1) and the Royal Society. X.-M. Jin acknowledges support from the National Young 1000 Talents Plan.

\bigskip

\section*{Supplementary Note 1 - Calculation of the hitting efficiency for classical random walks}

To calculate the hitting efficiency for classical random walks, we use the formalism of the versatile quantum stochastic walk algorithms [1-2]. In this way, even if here we focus on an extremal case (purely classical evolution), we leave open the possibility of extending the analysis to a more general scenario. The evolution of the density matrix $\rho$ describing the state of our walker can be evaluated by solving the Lindblad master equation

$$\frac{d\rho}{dt}=-(1-\omega)i[H,\rho]+\omega\sum_{ij}(L_{ij} \rho L_{ij}^\dagger- \frac{1}{2}\{L_{ij}^\dagger L_{ij}, \rho\}).\eqno{(S1)}$$  

The first part right to the equation represents the quantum walk evolution, where $H$ is the Hamiltonian operator. The second part, which contains the Lindblad operators $L_{ij}$, describes the classical random walk evolution. The parameter $\omega$ interpolates the weight of quantum walk and classical walk in the mixed quantum stochastic walk. When $\omega$ is 0, the evolution is purely quantum, and it turns to purely classical when $\omega$ becomes 1. This equation has been widely used for the study of open quantum systems, where classical interactions with the environment are present. We set $\omega$ to 1 to obtain the classical hitting efficiency dynamics on our graphs of different complexities. 

It is worth noting that the classical random walk slowly converges to a stationary result, where all the nodes have the same population. This population is thus equal to the inverse of the number of nodes in the graph, and we denote it as the average value $P_a$. In our investigation, we derive the classical hitting time by finding the time when the probabability of each waveguide has a deviation from $P_a$ of no more than $10^{-4}P_a$. In Fig.4.{\bf (b)}, the error bars for the classical hitting scenarios correspond to a range of criteria for judging the convergence: the upper bound corresponds to a deviation threshold of $10^{-5}P_a$ and the lower bound to a deviation threshold of $10^{-3}P_a$.  

\section*{Supplementary Note 2 - A comparison of experimental results with laser beam and heralded single photons}
The evolution pattern of a single quantum walker can be obtained by using either single photons or coherent light. We experimentally investigate the two approaches that are theoretically equivalent. We consider here a simpler setup: we use both a laser beam and a heralded single photon source to evaluate a one-dimensional quantum walk with different evolution lengths. The experimental evolution patterns are presented in Fig.5 with very high similarity. A small discrepancy occurs when the evolution length gets longer [see Fig.5.{\bf(e)} and Fig.5.{\bf(f)}]. The measured variance against the evolution length in double-logarithmic axes has a slope of 1.98 and 1.95 for laser beam and heralded single photons, respectively. This can be attributed to the differential bandwidth. The laser beam has a bandwidth of less than 1nm, while our heralded single-photon source is around 3nm [3-4].

\section*{Supplementary Note 3 - Sample characterization to seek the optimal evolution length}
All the experimental results measured with laser beam injection for hexagonal graphs with 2-8 layers are presented in Fig.6 - Fig.12. By scanning through 9 samples of the layer depth, we can find the optimal evolution length that has an optimal hitting efficiency. The measured hitting efficiency curves from the experimental results have an overall good agreement with the theoretical predictions. 

\newpage
\begin{figure*}[ht!]
\includegraphics[width=0.8\textwidth]{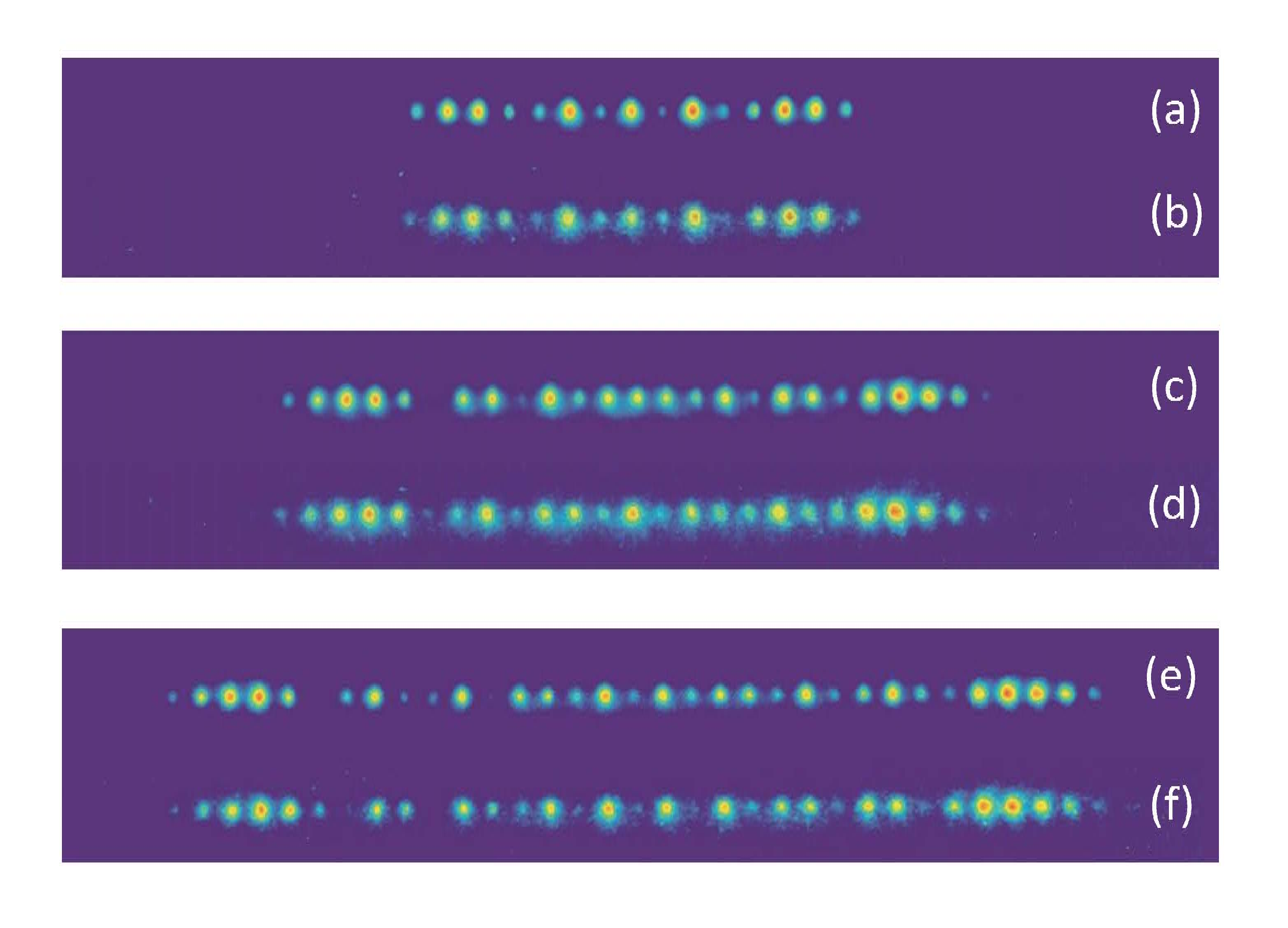}
\caption{\textbf{A comparison of laser beam and single photon measurements.} Measured population of the sites in a one-dimensional quantum walk with different evolution lengths using ({\bf a,c,e}) a laser beam and ({\bf b,d,f}) a single photon source, both vertically polarized and at a wavelength of 780nm. The evolution length is 1.5cm for ({\bf a,b}), 2.5cm for ({\bf c,d}), and 3.5cm for ({\bf e,f}).}
\end{figure*}

\begin{figure*}[ht!]
\includegraphics[width=0.8\textwidth]{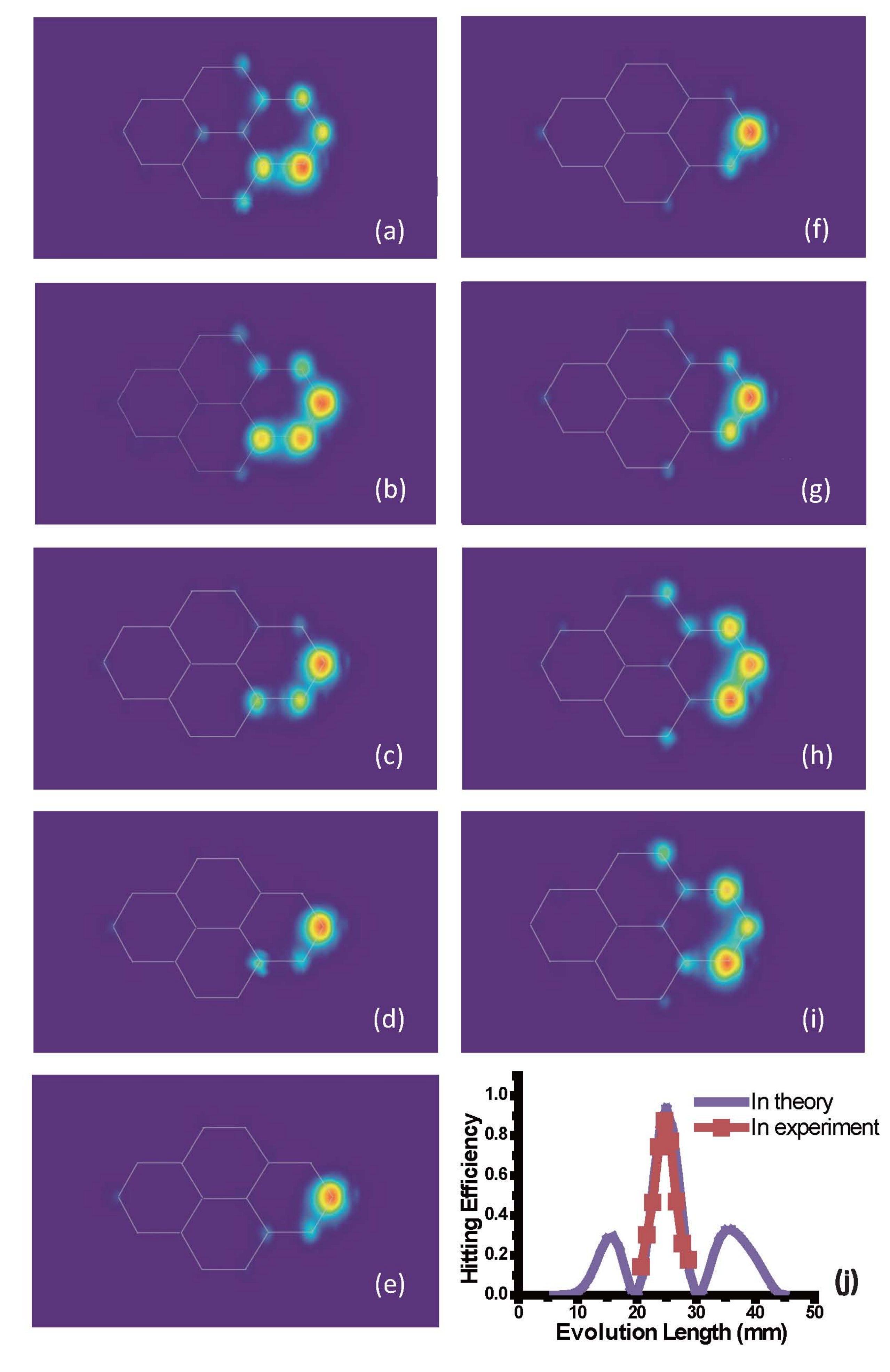}
\caption{\textbf{Experimental results for quantum hitting on a 2-layered hexagonal graph.} ({\bf a-i})  The photographed patterns for different evolution lengths: ({\bf a}) 20.7mm, ({\bf b}) 21.7mm, ({\bf c}) 22.7mm, ({\bf d}) 23.7mm, ({\bf e}) 24.7mm, ({\bf f}) 25.7mm, ({\bf g}) 26.7mm, ({\bf h}) 27.7mm, and ({\bf i}) 28.7mm.  ({\bf j}) The hitting efficiency against the evolution length for quantum hitting in experiment and theory.}
\end{figure*}

\begin{figure*}[ht!]
\includegraphics[width=0.8\textwidth]{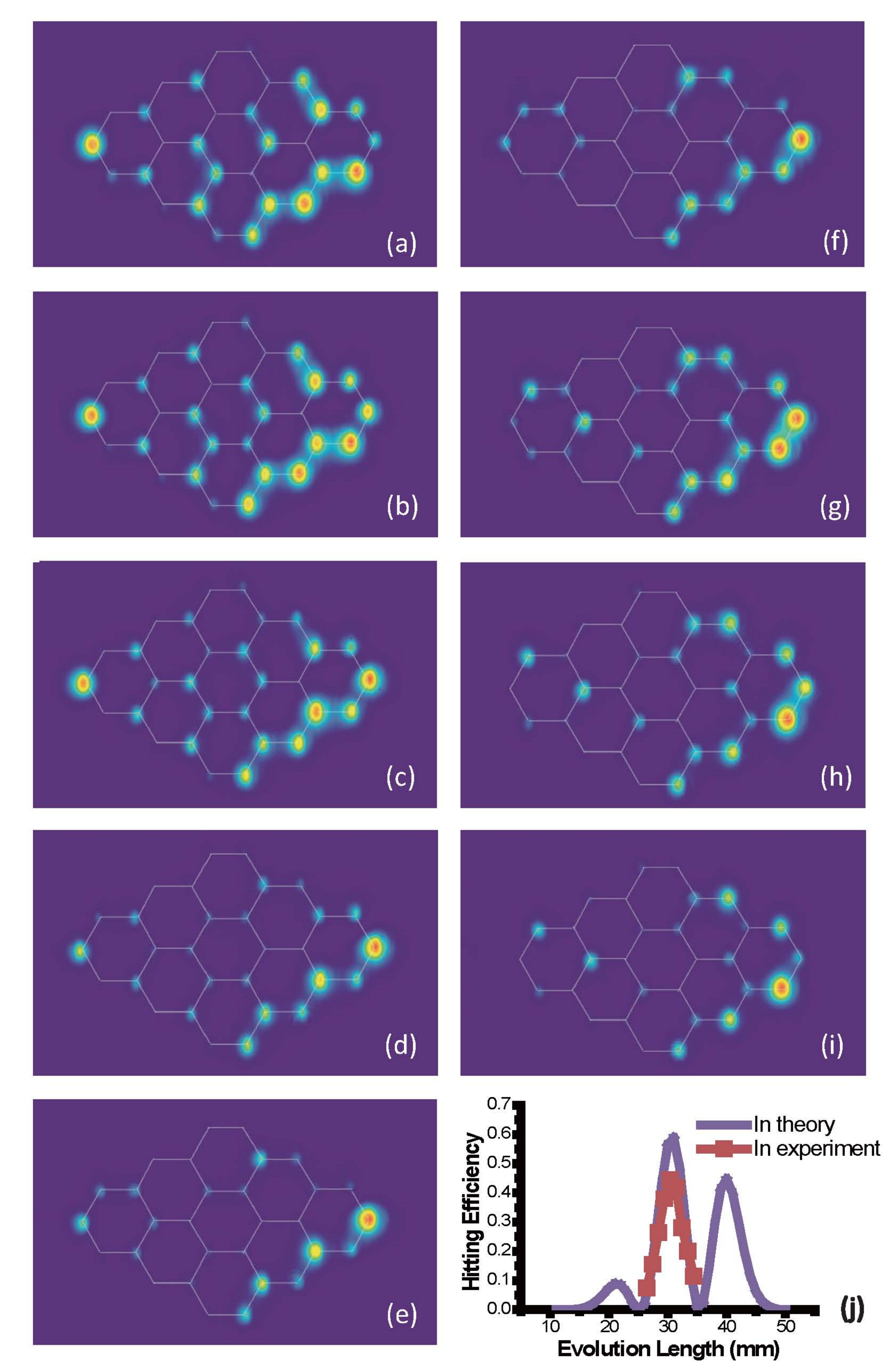}
\caption{\textbf{Experimental results for quantum hitting on a 3-layered hexagonal graph.} ({\bf a-i})  The photographed patterns for different evolution lengths: ({\bf a}) 26.4mm, ({\bf b}) 27.4mm, ({\bf c}) 28.4mm, ({\bf d}) 29.4mm, ({\bf e}) 30.4mm, ({\bf f}) 31.4mm, ({\bf g}) 32.4mm, ({\bf h}) 33.4mm, and ({\bf i}) 34.4mm.  ({\bf j}) The hitting efficiency against the evolution length for quantum hitting in experiment and theory.}
\end{figure*}

\begin{figure*}[ht!]
\includegraphics[width=0.8\textwidth]{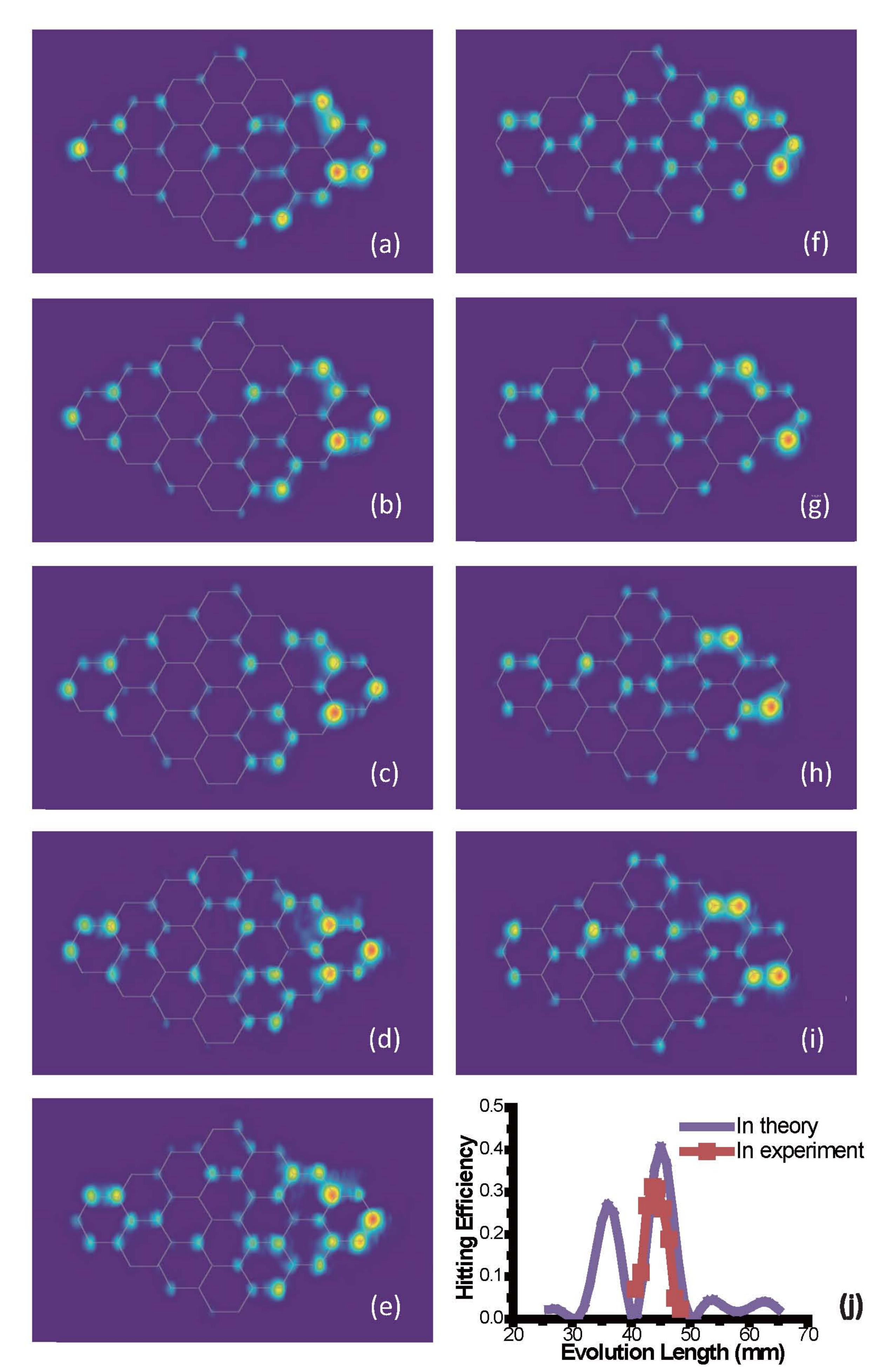}
\caption{\textbf{Experimental results for quantum hitting on a 4-layered hexagonal graph.} ({\bf a-i})  The photographed patterns for different evolution lengths: ({\bf a}) 40.7mm, ({\bf b}) 41.7mm, ({\bf c}) 42.7mm, ({\bf d}) 43.7mm, ({\bf e}) 44.4mm, ({\bf f}) 45.4mm, ({\bf g}) 46.4mm, ({\bf h}) 47.4mm, and ({\bf i}) 48.4mm.  ({\bf j}) The hitting efficiency against the evolution length for quantum hitting in experiment and theory.}
\end{figure*}

\begin{figure*}[ht!]
\includegraphics[width=0.8\textwidth]{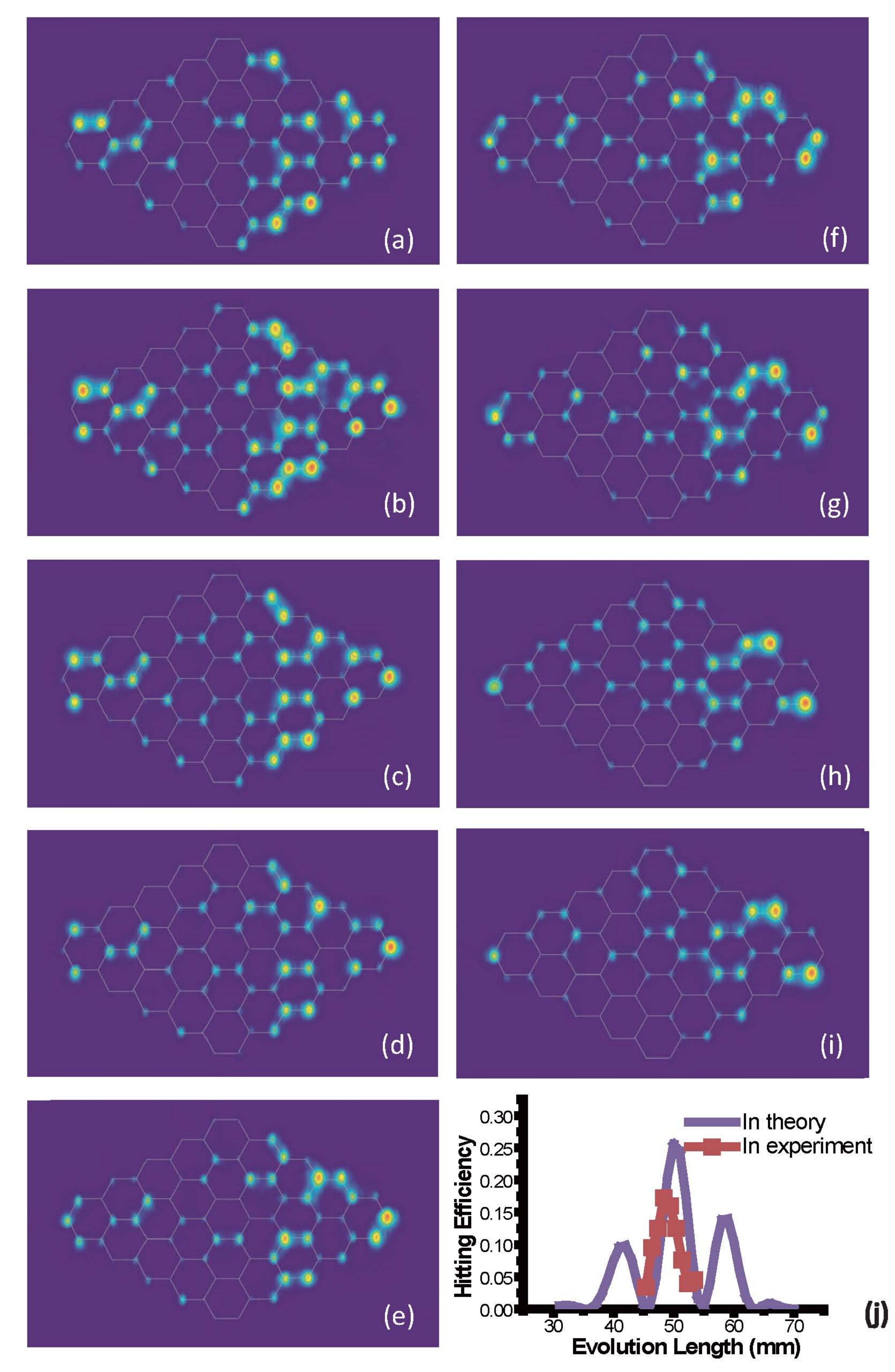}
\caption{\textbf{Experimental results for quantum hitting on a 5-layered hexagonal graph.} ({\bf a-i})  The photographed patterns for different evolution lengths: ({\bf a}) 45.4mm, ({\bf b}) 46.4mm, ({\bf c}) 47.4mm, ({\bf d}) 48.4mm, ({\bf e}) 49.4mm, ({\bf f}) 50.4mm, ({\bf g}) 51.4mm, ({\bf h}) 52.4mm, and ({\bf i}) 53.4mm.  ({\bf j}) The hitting efficiency against the evolution length for quantum hitting in experiment and theory.}
\end{figure*}

\begin{figure*}[ht!]
\includegraphics[width=0.8\textwidth]{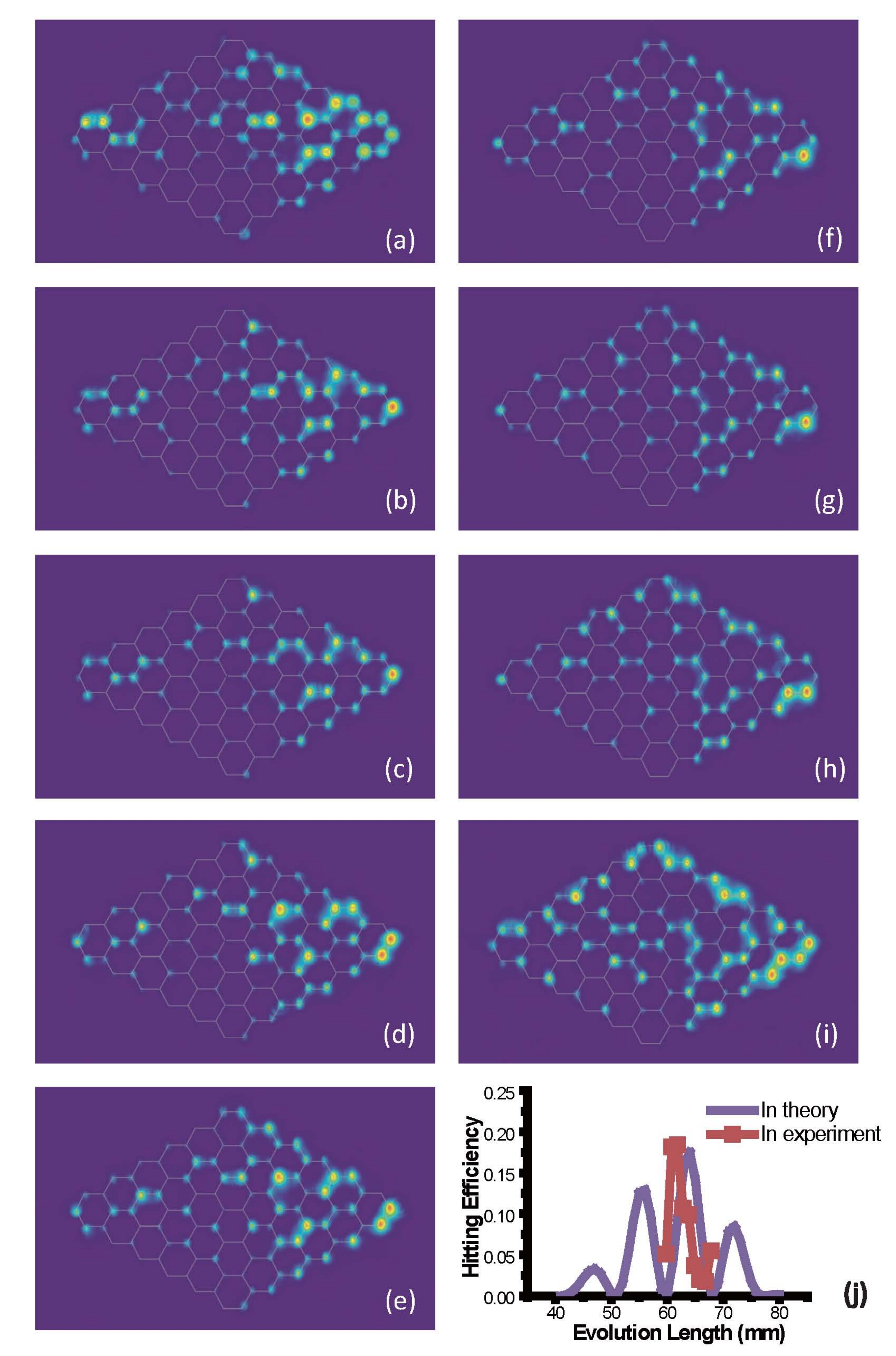}
\caption{\textbf{Experimental results for quantum hitting on a 6-layered hexagonal graph.} ({\bf a-i})  The photographed patterns for different evolution lengths: ({\bf a}) 59.8mm, ({\bf b}) 60.8mm, ({\bf c}) 61.8mm, ({\bf d}) 62.8mm, ({\bf e}) 63.8mm, ({\bf f}) 64.8mm, ({\bf g}) 65.8mm, ({\bf h}) 66.8mm, and ({\bf i}) 67.8mm.  ({\bf j}) The hitting efficiency against the evolution length for quantum hitting in experiment and theory.}
\end{figure*}

\begin{figure*}[ht!]
\includegraphics[width=0.8\textwidth]{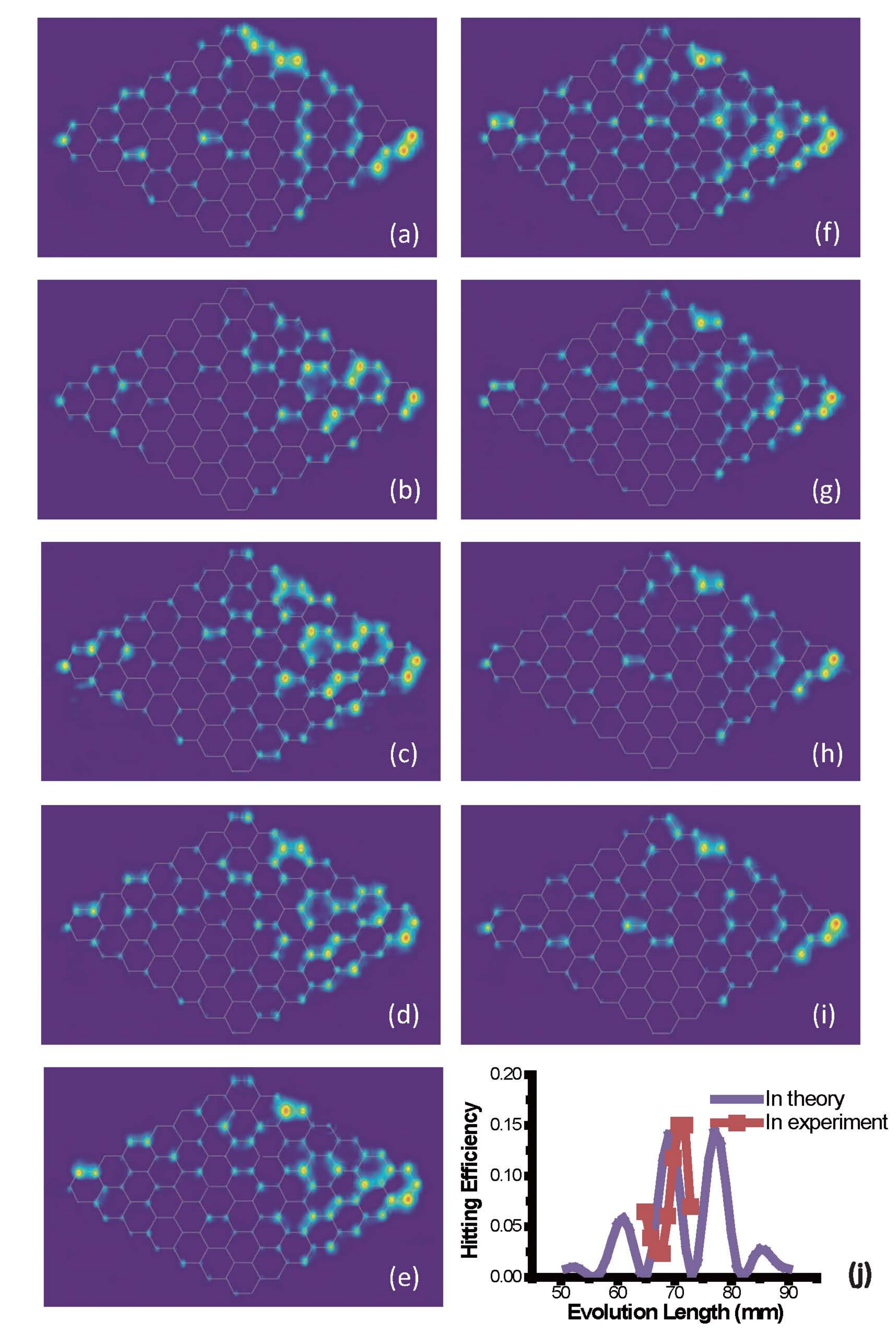}
\caption{\textbf{Experimental results for quantum hitting on a 7-layered hexagonal graph.} ({\bf a-i})  The photographed patterns for different evolution lengths: ({\bf a}) 64.8mm, ({\bf b}) 65.8mm, ({\bf c}) 66.8mm, ({\bf d}) 67.8mm, ({\bf e}) 68.8mm, ({\bf f}) 69.8mm, ({\bf g}) 70.8mm, ({\bf h}) 71.8mm, and ({\bf i}) 72.8mm.  ({\bf j}) The hitting efficiency against the evolution length for quantum hitting in experiment and theory.}
\end{figure*}

\begin{figure*}[ht!]
\includegraphics[width=0.8\textwidth]{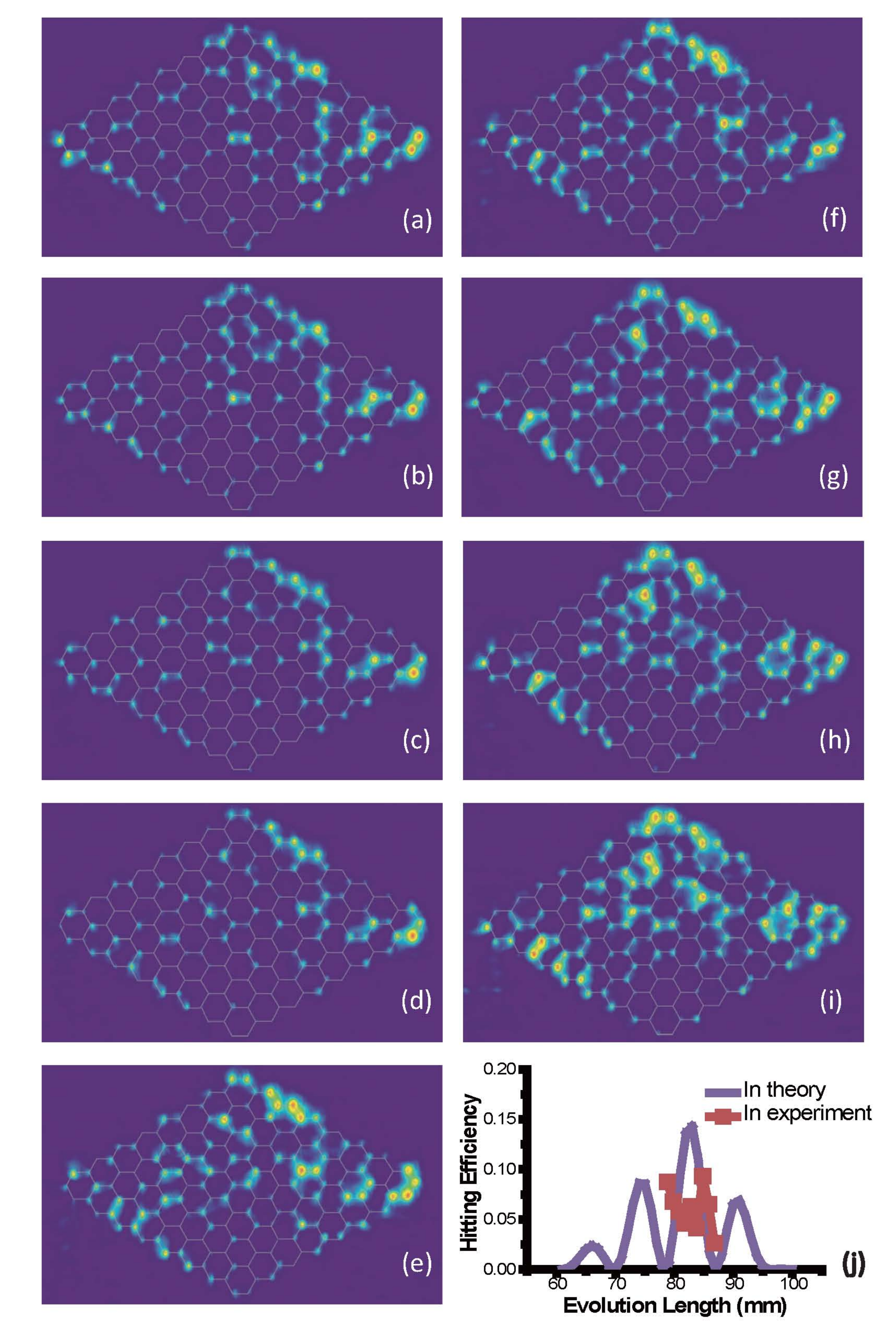}
\caption{\textbf{Experimental results for quantum hitting on a 8-layered hexagonal graph.} ({\bf a-i})  The photographed patterns for different evolution lengths: ({\bf a}) 78.8mm, ({\bf b}) 79.8mm, ({\bf c}) 80.8mm, ({\bf d}) 81.8mm, ({\bf e}) 82.8mm, ({\bf f}) 83.8mm, ({\bf g}) 84.8mm, ({\bf h}) 85.8mm, and ({\bf i}) 86.8mm.  ({\bf j}) The hitting efficiency against the evolution length for quantum hitting in experiment and theory.}
\end{figure*}

\newpage
\section*{Supplementary References}


\noindent [S1] Whitfield, J. D., Rodr\'iguez-Rosario, C. A., \& Aspuru-Guzik, A. Quantum stochastic walks: A generalization of classical random walks and quantum walks. \emph{Phys. Rev. Lett}. {\bf 81}, 022323 (2010).

\noindent [S2] Falloon, P. E., Rodriguez, J., \& Wang, J. B. QSWalk: A \emph{Mathematica} package for quantum stochastic walks on arbitrary graphs. \emph{Comput. Phys. Commun.} {\bf 217}, 162-170 (2017).

\noindent [S3] Kim, Y. H., Kulik, S. P., Chekhova, M. V., Grice, W. P., \& Shih, Y. H. Experimental entanglement concentration and universal Bell-state synthesizer. \emph{Phys. Rev. A} {\bf 67}, 010301(R) (2003).

\noindent [S4] Kim, Y. H. Quantum interference with beamlike type-II spontaneous parametric down-conversion. \emph{Phys. Rev. A} {\bf 68}, 013804 (2003).

\end{document}